\documentclass[aps,pra,twocolumn,showpacs]{revtex4}
\input{epsf}
\usepackage{epsfig}
\begin{document}

\title{Coherent generation of EPR-entangled light pulses
mediated by a single trapped atom}

\author{Giovanna Morigi}
\affiliation{Grup d'Optica, Departament de Fisica, Universitat
Autonoma de Barcelona, 08193 Bellaterra, Spain}
\author{J\"urgen Eschner}
\affiliation{ICFO - Institut de Ci\`{e}ncies Fot\`{o}niques, 08860
Castelldefels (Barcelona), Spain}
\author{Stefano Mancini}
\author{David Vitali}
\affiliation{CNISM and Dipartimento di Fisica, Universit\`a di
Camerino, 62032 Camerino, Italy}

\date{\today}

\begin{abstract}
We show that a single, trapped, laser-driven atom in a
high-finesse optical cavity allows for the quantum-coherent
generation of entangled light pulses on demand. Schemes for
generating simultaneous and temporally separated pulse pairs are
proposed. The mechanical effect of the laser excitation on the
quantum motion of the cold trapped atom mediates the entangling interaction
between two cavity modes and between the two subsequent pulses,
respectively. The entanglement is of EPR-type, and its degree can
be controlled through external parameters. At the end of the
generation process the atom is decorrelated from the light field.
Possible experimental implementations of the proposals are
discussed.
\end{abstract}


\maketitle

\section{Introduction}

One of the most intriguing features of quantum mechanics is the
possibility of entangling physical systems, which has both
fundamental and practical implications. In particular,
entanglement has been recognized as a valuable resource for
quantum information processing and for cryptography. In this
context, two approaches have been developed, one based on discrete
variables, the other using continuous variables. The main
motivation to deal with continuous variables originates from
practical considerations: efficient implementation of the
essential steps of quantum information processing are achievable
in quantum optics, utilizing the continuous quadrature variables
of the quantized electromagnetic field~\cite{Bra04}. In the
continuous variable setting, Gaussian states play a prominent
role, and when considering bipartite Gaussian systems, entangled
states are synonymous to two-mode squeezed states~\cite{Bra04},
and their entanglement is equivalent to the position-momentum
entanglement originally considered by Einstein, Podolsky, and
Rosen (EPR)~\cite{Reid}.

Conventionally, two-mode squeezed states emerge from the nonlinear
optical interaction of a laser with a crystal, i.e.\ from
parametric amplification or oscillation. As such, the phenomenon
is the result of many-atom dynamics (often described by a simple
nonlinear polarizability model). Then, an interesting question is
whether analogous macroscopic nonlinear phenomena can emerge as
well from the quantum dynamics of a \emph{single} atom.

Recently, several experimental realizations have accessed novel
regimes of engineering atom-photon interaction and opened
promising perspectives for implementing controlled nonlinear
dynamics with simple quantum optical systems. Examples are
entangled light generation in atomic ensembles \cite{Kimble03,
Giacobino}, atomic memory for quantum states of light
\cite{LukinScience03, Lukin03, Polzik04}, entanglement between a single atom and its
emitted photon \cite{Monroe04, Weinfurter2005}, entanglement of
remote ensembles \cite{Chou2005}, one-atom laser \cite{An94,
Kimble-atomlaser}, mechanical forces of single photons on single
atoms \cite{Kimble-fly, Rempe-fly, Bushev2005}, controlled
interaction of a trapped ion and a cavity \cite{Guthoehrlein01,
Mundt02}, controlled single-photon generation \cite{Kuhn02,
Kimble-photon, Keller04}, as well as quantum state and
entanglement engineering in the microwave regime \cite{MicroCQED}.

\begin{center}
\begin{figure}[htb]
\epsfig{width=0.99\hsize, file=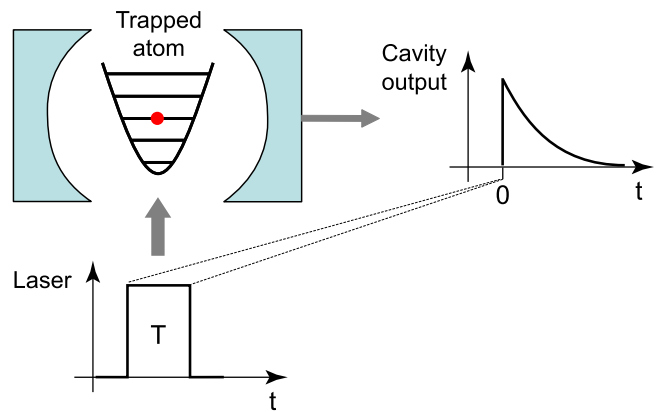} %
\caption{A trapped atom (ion) is confined inside a good resonator and is driven by a
laser pulse of duration $T$ which propagates orthogonally to the cavity axis. By coupling the external
atomic degrees of freedom to the cavity dynamics the pulse prepares the cavity field in a
non-classical state, which is transmitted to the output by cavity decay. A basic condition for the validity of these dynamics is $\kappa T\ll 1$, whereby $\kappa$ is the cavity decay rate.} \label{Fig:0}
\end{figure}
\end{center}

In this work we investigate the realization of an optical
parametric amplifier based on a single cold trapped atom inside a
high-finesse optical cavity and driven by a short laser pulse, as sketched
in Fig.~\ref{Fig:0}. We show theoretically that this
system allows for the controlled, quantum-coherent generation of
entangled light pulses by exploiting the mechanical effects of
atom-photon interaction. The pulses can contain many ($\gg 1$)
photons, and their entanglement is of continuous-variable (or EPR)
type.

We present the detailed study of two schemes, which have been
proposed in~\cite{PRL}, and discuss in particular experimental
conditions for their realization.
The first scheme requires a bichromatic cavity interacting with the atomic dipole.
Here, we show that--after short coherent excitation of the atom by an external laser
beam--the cavity emits a bichromatic pulse of two-mode squeezed,
i.e.\ entangled light. The second scheme relies on the interaction between the dipole and one cavity mode.
Here, creation of two subsequent, entangled pulses at the cavity output is
accomplished by using two temporally separated excitation pulses, with the quantum state
of the atomic motion serving as intermediate memory which mediates the entanglement between the
first and second pulse at the cavity output.
In both cases entanglement is found on time scales of the
order of the cavity decay time. In particular, noise reduction
below 10\% of the shot-noise level in the relative amplitude
fluctuations of the two light fields is derived for an
experimentally accessible set of parameters. Variation of the
coupling parameters between atom and light allows for tuning the
degree of entanglement between the cavity modes. Moreover, the
emitted light pulses are decorrelated from the atom, i.e.\ at the end
of the process the atom carries no memory of the interaction.

Our scheme applies concepts developed for macroscopic
oscillators~\cite{Mancini02,Pirandola03} to a single quantum optical system,
by exploiting the coupling between internal (electronic) and
external (motional) degrees of freedom. Such coupling is
negligible in macroscopic systems, but significant in atomic
systems, thus rendering the dynamics far more accessible. Our
study is also connected to ideas of mapping quantum states of
atoms onto light inside a cavity~\cite{Ze-Parkins99}, to their
implementation for establishing entanglement between distant
atoms~\cite{Parkins02}, and to recent experimental and theoretical
studies on quantum correlations in the light scattered by
atoms~\cite{LukinScience03, Kimble03, Polzik04, Giacobino,
Polaritons,Jakob99,NaGutKike}. The proposal differs fundamentally from
existing methods for generating pulsed squeezing \cite{kumar} or
intense pulses of polarization-entangled photons
\cite{Bouwmeester} which employ nonlinear crystals driven by a
pulsed pump: in our case the microscopic nature of the medium
allows for full coherent control of the light-matter quantum
correlations and of the final quantum state of the generated
light.

The paper is organized as follows. In Sec.~\ref{Sec:Theory} we
define the master equation governing the coherent and dissipative
dynamics of the internal and motional degrees of freedom of the
laser-driven atom and of the two relevant modes of the cavity
coupling to the atom. In Sec.~\ref{Sec:H_eff} we derive the approximate Hamiltonian
effecting the dynamics which lead to the generation of
entanglement between the cavity modes, and in Sec.~\ref{Sec:Dynamics} we discuss
the dynamics it generates. In Sec.~\ref{Sec:Output} we introduce the field at the cavity output and in Sec.~\ref{Sec:Entanglement} we investigate its quantum correlations. Section~\ref{Sec:Experiment} discusses the possibility of experimentally realizing the proposed scheme,
i.e., the required experimental setup and parameters which
would allow to observe the dynamics. Then, in
Sec.~\ref{Sec:Time_separation} we discuss the (conceptually
simpler) case of creating temporally separated, entangled pulses
in one and the same cavity mode. The conclusions are drawn in
Sec.~\ref{Sec:Conclusions}, and in the appendix details of the
calculations at the basis of the results in
Sec.~\ref{Sec:Entanglement} are provided.

\section{Simultaneous bichromatic pulses}
\label{Sec:II}

In this section we present a scheme for the simultaneous generation of bichromatic entangled light pulses. The scheme bases itself on the interaction between a bimodal cavity and the dipole of a trapped atom, which is driven by a laser pulse. EPR-type entanglement is established between the cavity modes via the quantum motion of the center-of-mass. As a result the light pulses emitted at the cavity output exhibit quantum correlations of EPR-type.

\subsection{Theoretical model} \label{Sec:Theory}

We consider an atom of mass $m$, which is confined inside an
optical cavity by an external potential, as shown in Fig.~\ref{Fig:0}.
The center-of-mass motion is along the $\hat{x}$ axis, as we assume that the radial
potential is sufficiently steep, that the motion in this plane can
be considered frozen out. The potential along $\hat{x}$ is
harmonic with frequency $\nu$. Position and momentum of the atomic
center-of-mass are denoted by $x$ and $p$, respectively. The
corresponding center-of-mass dynamics are given by
\begin{equation}
\label{H:mec} H_{\rm
mec}=\frac{p^2}{2m}+\frac{1}{2}m\nu^2x^2=\hbar\nu\left(b^{\dagger}b+\frac{1}{2}\right)
\end{equation}
where $b,b^{\dagger}$ are the annihilation and creation operators,
respectively, of a quantum of vibrational energy $\hbar\nu$, with
$x=\sqrt{\hbar/2m\nu}(b+b^{\dagger})$ and $p={\rm i}\sqrt{\hbar
m\nu/2}(b^{\dagger}-b)$. We denote by $|n_{\rm mec}\rangle$ the
eigenstates of $H_{\rm mec}$ at energy $\hbar\nu (n_{\rm
mec}+1/2)$. The atom's relevant internal degrees of freedom are described by a
ground state $|g\rangle$ and an excited state $|e\rangle$ which
form a dipole with dipole moment ${\bf d}$ and frequency
$\omega_0$, and the atomic Hamiltonian has the form
\begin{equation}
H_a=\hbar\omega_0|e\rangle\langle e|+H_{\rm mec}~.
\label{H:atom}
\end{equation}
The full dynamics are described by Hamiltonian
$$H=H_a+H_c+H_{ac}+H_{aL}~,$$
where the terms $H_c$, $H_{ac}$ and $H_{aL}$ describe two cavity modes and the
coupling of the atomic dipole to the electromagnetic field of the cavity
and of a laser, respectively. The cavity Hamiltonian is
\begin{equation}
H_c=\sum_{j=1,2}\hbar\omega_ja_j^{\dagger}a_j~,
\end{equation}
where $\omega_j$ are the frequencies of two optical modes, and
$a_j,a_j^{\dagger}$ are the respective annihilation and
creation operators of a quantum of energy $\hbar\omega_j$, i.e.\ a
photon in mode $j$. We denote by $|n_1,n_2\rangle$ the eigenstates of $H_c$ at energy
$\hbar\omega_1n_1+\hbar\omega_2n_2$. The coupling between the dipole and the cavity
modes is represented by
\begin{eqnarray}
\label{Hint} H_{ac} =\hbar\sum_{j=1,2} g_j
a_j\sigma^{\dagger}\cos(k_jx\cos\theta_c+\phi_j) + {\rm H.c.}
\end{eqnarray}
whereby the modes have wave vectors $\vec{k}_j$
($k_j=|\vec{k}_j|$) forming an angle $\theta_c$ with the axis
$\hat{x}$ of the motion, and $g_j$ is the coupling strength of the
dipole to the corresponding mode. The angle $\phi_j$ takes into
account the position of the trap center inside the cavity. The
terms $\sigma = |g\rangle\langle e|$ and $\sigma^{\dagger} =
|e\rangle\langle g|$ denote the dipole lowering and raising
operators. The coupling to the laser at frequency $\omega_L$ is
\begin{equation} \label{HL}
H_{aL}=\hbar\Omega(t) \sigma^{\dagger}{\rm e}^{-{\rm i}(\omega_L
t-k_Lx\cos\theta_L)}+{\rm H.c.}~,
\end{equation}
where $\Omega$ is the (slowly varying) Rabi frequency
and $\vec{k}_L$ is the wave vector ($k_L=|\vec{k}_L|$) forming an angle
$\theta_L$ with the trap axis.
In what follows we drop the subscripts in the moduli of the
wave vectors and denote them by $k$.

Denoting by $\rho$ the density matrix of the cavity modes and of
the atom's internal and external degrees of freedom, the master
equation for the dynamics reads
\begin{equation}
\label{M:Eq} \frac{\partial}{\partial t}\rho=\frac{1}{{\rm
i}\hbar}[H,\rho]+{\cal K}\rho+{\cal L}\rho
\end{equation}
which accounts for the coherent interaction of the dipole with the
cavity modes, and the incoherent processes constituted by
spontaneous emission and cavity decay, namely
\begin{eqnarray}
&&{\cal
L}\rho=\frac{\gamma}{2}\left(2\sigma\tilde{\rho}\sigma^{\dagger}-\sigma^{
\dagger}\sigma\rho
-\rho\sigma^{\dagger}\sigma\right)\\
&&{\cal K}\rho=\sum_{j=1,2}\kappa_j\left(2a_j\rho
a_j^{\dagger}-a_j^{\dagger}a_j\rho -\rho
a_j^{\dagger}a_j\right)\label{K}
\end{eqnarray}
where $\gamma$ is the spontaneous emission rate of the atom into
modes external to the cavity and $\kappa_j$ are the decay rates of
the cavity modes. The density matrix $\tilde{\rho}$ accounts for
the mechanical effect of photon emission,
\begin{eqnarray*}
\tilde{\rho}=\int_{-1}^{1}{\rm d}u{\cal N}(u){\rm e}^{-{\rm
i}kux}\rho~{\rm e}^{ {\rm i}kux}
\end{eqnarray*}
with probability ${\cal N}(u){\rm d}u$ that the spontaneously
emitted photon imparts a recoil momentum $\hbar k u$ to the atom.

We assume that the atomic motion is in the Lamb-Dicke regime, and
expand the interaction terms~(\ref{Hint}) and~(\ref{HL}) to second
order in the Lamb-Dicke parameter $\eta=k\sqrt{ \hbar/2m\nu}$. In
this limit they take the form
\begin{eqnarray}
\label{Hint:2} H_{ac} &=& \hbar\sum_{j=1,2}
g_j\cos\phi_j\Bigl(a_j\sigma^{\dagger}\bigl(1-\eta\cos\theta_c\tan\phi_j(b^{\dagger}+b)
\nonumber\\
& &-\frac{\eta^2}{2}\cos^2\theta_c(b^{\dagger}+b)^2\bigr)+{\rm
O}(\eta^3)\Bigr)+{\rm H.c.}
\end{eqnarray}
and
\begin{eqnarray}
\label{HL:2} H_{aL} &=&\hbar\Omega(t) \sigma^{\dagger}{\rm
e}^{-{\rm i}\omega_L t}\Bigl(1+{\rm
    i}\eta\cos\theta_L(b^{\dagger}+b) \\
& &-\frac{\eta^2}{2}\cos^2\theta_L(b^{\dagger}+b)^2\bigr)+{\rm
O}(\eta^3)\Bigr) +{\rm H.c.}\nonumber
\end{eqnarray}

\begin{center}
\begin{figure}[htb]
\epsfig{width=0.99\hsize, file=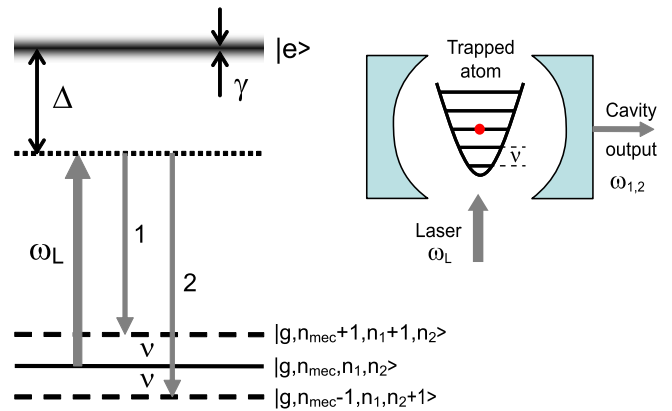} %
\caption{Layout of the system and energy diagram. A single atom
with internal energy levels $|g\rangle$ and $|e\rangle$ is
confined by an external potential inside an optical cavity and is
driven by a laser. The orientation of the considered vibrational mode
has non-zero projection onto the laser direction.
The harmonic motion, at frequency $\nu$, modulates the laser frequency, $\omega_L$, and
the Stokes and anti-Stokes components at $\omega_L \pm \nu$ are
resonant with two cavity modes, labeled 1 and 2. The linewidth of
$|e\rangle$ is $\gamma$, and $\Delta$ is the detuning between
laser and atom; $n_{\rm mec},n_1,n_2$ label the number of excitations of the center-of-mass, cavity mode 1 and cavity mode 2 oscillator, respectively.} \label{Fig:1}
\end{figure}
\end{center}

\subsection{Effective Hamiltonian}
\label{Sec:H_eff}

We consider the reference frame rotating at the laser frequency,
and denote by
$$\Delta=\omega_{\rm L}-\omega_0$$ the detuning between laser and atom, and by
$$\delta_j=\omega_{\rm L}-\omega_j$$ the detunings between the cavity
modes and the laser. In particular, we assume
\begin{eqnarray*}
\delta_1=\nu~~;~~\delta_2=-\nu
\end{eqnarray*}
namely, the mode frequencies are spaced by the quantity $2\nu$,
and the laser frequency is tuned symmetrically between them.
Hence, in this reference frame
$$H_{\rm c}^{\prime}=\hbar\nu (a_1^{\dagger}a_1-a_2^{\dagger}a_2)~.$$
This choice of the
frequency spacing and detunings allows us to select certain
resonant scattering processes which determine the dynamics on the
considered time scales.
Figure~\ref{Fig:1} dysplays the quantum states which are resonantly coupled for this choice of the parameters.
The corresponding effective Hamiltonian is derived in the following.

In the limit $$|\Delta|\gg \Omega,g_j,\gamma$$ we eliminate the
atom's internal degrees of freedom in second order perturbation
theory in the parameter $\Omega/|\Delta|$ and obtain the
approximate Hamiltonian
\begin{equation}
\label{H:eff} H_{\rm eff}=H_{\rm c}^{\prime}+H_{\rm mec}+H_1+H_2
\end{equation}
which is defined on the subspace $|g,n_{\rm mec},n_1,n_2\rangle$. The dynamics of
the coupling are given by the terms
\begin{eqnarray}
&&H_1={\rm i}\hbar\chi_1 a_1^{\dagger}b^{\dagger}+{\rm
 H.c}~,\label{H:1}\\
&&H_2={\rm i}\hbar\chi_2a_2^{\dagger}b+{\rm H.c}~,\label{H:2}
\end{eqnarray}
with
\begin{eqnarray}
&&\chi_1=\eta
g_1^*\cos\phi_1\Omega(t)\left(\frac{\cos\theta_L}{\Delta-\nu+{\rm i}\gamma/2}
+\frac{{\rm
i}\tan\phi_1\cos\theta_c}{\Delta+{\rm i}\gamma/2}\right)~,\nonumber\\
&&\label{Chi:1}\\
&&\chi_2=\eta
g_2^*\cos\phi_2\Omega(t)\left(\frac{\cos\theta_L}{\Delta+\nu+{\rm i}\gamma/2}
+\frac{{\rm i}\tan\phi_2\cos\theta_c}{\Delta+{\rm i}\gamma/2}\right)~.\nonumber\\
&&\label{Chi:2}
\end{eqnarray}
Hamiltonian~(\ref{H:1}) describes an interaction giving rise to
two-mode squeezing between the center-of-mass oscillator and the
cavity mode at frequency $\omega_1$.
Hamiltonian~(\ref{H:2}) describes a beam-splitter
type of interaction between the center-of-mass oscillator and the
cavity mode at frequency $\omega_2$~\cite{Ze-Parkins99}. Their coupling strengths
$\chi_1,\chi_2$, Eqs.~(\ref{Chi:1},~\ref{Chi:2}), depend on the
value of the atom-cavity coupling constants $g_1$, $g_2$, on the
geometry of the setup, and on the ratio between the trap frequency
$\nu$ and the laser detuning $\Delta$. In particular, each of them
is the sum of two terms, which represent two indistinguishable
paths leading to the creation of a cavity photon accompanied by
the creation ($\chi_1$) or annihilation ($\chi_2$) of a
vibrational quantum. This interference depends on the geometry of
the setup and may lead to significantly different values of
$\chi_1$ and $\chi_2$ when the ratio $\nu/|\Delta|$ is not too
small~\cite{Cirac93,Zippilli05}.

It should be remarked that these equations have been obtained at
first order in the Lamb-Dicke expansion, neglecting off-resonant
and inelastic scattering processes. They are valid on a time scale
in which the resonant processes, in which a photon is scattered
into the cavity mode under annihilation or creation of a
vibrational quantum, dominate over all other processes. In
addition, we have assumed that during these dynamics the cavity
does not decay. This assumption implies that the duration of the
laser pulses $T$ is much shorter than the cavity lifetime, $\kappa
T\ll 1$. On the other hand, the dynamics are based on the spectral
resolution of the cavity modes spaced by twice the trap frequency,
i.e.\ $\nu T\gg 1$. Therefore, relation
\begin{equation}
\label{Condition} \kappa\ll \frac{1}{T}\ll \nu
\end{equation}
is required for the validity of the equations derived above. We
refer the reader to Sec.~\ref{Sec:Experiment} for an extensive
discussion of the parameter regimes in which
Hamiltonian~(\ref{H:eff}) holds. Note that the photons which are elastically
scattered into modes external to the cavity do not affect the
center-of-mass or cavity mode dynamics. Therefore, they can be
traced out from the respective equations of motion without causing
decoherence.

\subsection{Dynamics}
\label{Sec:Dynamics}

Let us now discuss the coherent physical dynamics that
Eq.~(\ref{H:eff}) describes, and neglect for the moment incoherent processes.
In this case the observable $C = b^{\dagger}b -
a_1^{\dagger}a_1 + a_2^{\dagger}a_2$ is a constant of the motion.
Therefore, if we consider the state $|n_{\rm mec},0_1,0_2\rangle$
at $t=0$, it will be coupled to states of the type $|n_{\rm
mec}+l_1-m_2,l_1,m_2\rangle$, which are eigenstates of $C$ at the
same eigenvalue $n=n_{\rm mec}$. The Heisenberg equations of
motion
\begin{eqnarray}
\label{Heisen:1}
&&\dot{a}_1=\chi_1b^{\dagger}\\
&&\dot{b}=\chi_1a_1^{\dagger}-\chi_2^* a_2
\label{Heisen:2}\\
&&\dot{a}_{2}=\chi_2b\label{Heisen:3}
\end{eqnarray}
generate periodic dynamics provided that $|\chi_2|>|\chi_1|$. In
this case their solutions read~\cite{Pirandola03}
\begin{eqnarray}
\label{a_1} a_1(t)
&=&\frac{\chi_1}{\Theta}b^{\dagger}(0)\sin\Theta t+\frac{1}{\Theta^2}\left[|\chi_2|^2-|\chi_1|^2\cos\Theta t\right]a_1(0)\nonumber\\
& &-\frac{\chi_1\chi_2}{\Theta^2}\left[1-\cos\Theta t\right]a_2^{\dagger}(0)~,\\
a_2(t) &=&\frac{\chi_2}{\Theta}b(0)\sin\Theta t+
\frac{\chi_1\chi_2}{\Theta^2}\left[1-\cos\Theta t\right]a_1^{\dagger}(0)\nonumber\\
& &-\frac{1}{\Theta^2}\left[|\chi_1|^2-|\chi_2|^2\cos\Theta t\right]a_2(0)~,\\
b(t) &=&b(0)\cos\Theta
t+\frac{1}{\Theta}\left[-\chi_2^*a_2(0)+\chi_1 a_1^{\dagger}(0
)\right]\sin\Theta t~,\nonumber\\
\label{b}
\end{eqnarray}
with
\begin{equation}\Theta=\sqrt{|\chi_2|^2-|\chi _1|^2}~.
\end{equation}
In general these solutions describe tripartite entanglement among
cavity modes and center-of-mass oscillator~\cite{Pirandola03}. An
interesting situation is found after half a period, for
$T_{\pi}=\pi/\Theta$. At this time (modulus $2\pi$) we find
\begin{eqnarray}
&&a_1(T_{\pi})=\frac{|\chi_1|^2+|\chi_2|^2}{\Theta^2}a_1(0)-\frac{2\chi_1\chi_2}{\Theta^2}a_2^{\dagger}(0)~,\\
&&a_2(T_{\pi})=\frac{2\chi_1\chi_2}{\Theta^2}a_1^{\dagger}(0)-\frac{|\chi_1|^2+|\chi_2|^2}{\Theta^2}a_2(0)~,\\
&&b(T_{\pi})=-b(0)~.
\end{eqnarray}
Hence, at this instant the center-of-mass oscillator is
decorrelated from the cavity modes. For instance, if at $t=0$ the
center-of-mass oscillator density matrix is a thermal state at
temperature ${\cal T}$ given by
$$\mu(0)=(1-{\rm e}^{-\beta\hbar\nu}){\rm e}^{-\beta H_{\rm
mec}},$$ with $\beta=1/k_B{\cal T}$, then $\mu(T_{\pi})=\mu(0)$.
Most remarkably, however, if at $t=0$ the cavity modes are in the
vacuum, then at $t=T_{\pi}$ they exhibit EPR-type
entanglement~\cite{Reid}, their state being the two-mode squeezed
state
\begin{equation}
\label{Two:mode}
|\psi\rangle=\left(\frac{1-r^2}{1+r^2}\right)\sum_{n=0}^{\infty}\left[-\frac{2r}{1+r^2}e^{i\phi}\right]^n
|n,n\rangle,
\end{equation}
where
\begin{equation}
r=\left|\frac{\chi_2}{\chi_1}\right| \label{r}
\end{equation}
and $\phi=\arg(\chi_1)+\arg(\chi_2)$. The average number of
photons per mode is
\begin{equation}
\langle n\rangle=4r^2/(1-r^2)^2~.
\end{equation}
Hence, if the laser pulse has duration $T_{\pi}$, after the
interaction the cavity modes are EPR-entangled with each other and
decorrelated from the quantum state of the center-of-mass motion.
The mechanical effects of the atom-photon interaction plays a
fundamental role in establishing the entanglement, nevertheless
the initial motional state does not affect the efficiency of the
process.

\subsection{Field at the cavity output}
\label{Sec:Output}

In this section we introduce the theoretical description of the
field at the cavity output, which will be used in
Sec.~\ref{Sec:Entanglement} for determining the degree of quantum
correlation of the emitted pulses. The cavity output is described
by the Heisenberg operator~\cite{Carmichael}
\begin{equation}
{\bf E}(x',t)={\bf E}^{(+)}(x',t)+{\bf E}^{(-)}(x',t)
\end{equation}
where ${\bf E}^{(+)}(x',t)$ is the negative frequency part and
${\bf E}^{(-)}(x',t)$ its adjoint at the position $x'$ outside the
cavity, setting the mirror at $x'=0$. We decompose the field into
the free and the source field terms, according to
\begin{equation}
\label{E:s-f} {\bf E}^{(+)}(x',t)={\bf E_s}^{(+)}(x',t)+{\bf
E_f}^{(+)}(x',t)~,
\end{equation}
whereby the source field is given by
%\begin{eqnarray}
%{\bf E_s}^{(+)}(x',t) &=& {\rm i}\sum_j\hat{e}_j{\rm e}^{{\rm i}\phi_{Tj}}
%\sqrt{\frac{\hbar\omega_{j}}{2\epsilon_0Ac}} \\
%&\times& \sqrt{2\kappa_j}a_j(t-x'/c)~. \label{E:s} \nonumber
%\end{eqnarray}
\begin{eqnarray}
{\bf E_s}^{(+)}(x',t) &=& {\rm i}\sum_j \hat{e}_j{\rm e}^{{\rm
i}\phi_{Tj}} \sqrt{\frac{\hbar\omega_{j}}{2\epsilon_0Ac}}
\sqrt{2\kappa_j}a_j(t-x'/c)~, \nonumber\\
\label{E:s}
\end{eqnarray}
and only times $t>x'/c$ are considered. Here, $A$ is the
cross-sectional area of the cavity mode, and $\phi_{Tj}$ is the
phase change on transmission through the output mirror. The free
field is
\begin{eqnarray} {\bf E_f}^{(+)}(x',t)
={\rm
i}\sum_{j,k}\hat{e}_j\sqrt{\frac{\hbar\omega_{k}}{2\epsilon_0AL'}}
r_k^{(j)}(0){\rm e}^{-{\rm i}[\omega_k(t-x'/c)-\phi_R]}~, \nonumber\\
\label{E:f}
\end{eqnarray}
which is defined for $x'>0$. Here, $r_{k}^{(j)}$ and
$r_{k}^{(j)\dagger}$ are annihilation and creation operators for
the modes of the electromagnetic field external to the cavity at
frequency $\omega_k$ and polarization $\hat{e}_j$; $L'$ is the
quantization length at the cavity output, and $\phi_R$ the phase
change upon reflection at the cavity output mirror.

Using
\begin{eqnarray}
r_f^{(j)}(t)={\rm e}^{{\rm
i}(\phi_R-\phi_{Tj})}\sum_k\sqrt{\frac{\omega_k}{\omega_j}}
r_k^{(j)}(0){\rm e}^{-{\rm i}\omega_kt}~,
\end{eqnarray}
we introduce the rescaled field operator $Q_j^{(+)}(x',t)$ whereby
\begin{eqnarray} \label{Rescale}
Q_j^{(+)}(x',t) =
\sqrt{c/L'}r_f^{(j)}(t-x'/c)+\sqrt{2\kappa_j}a_j(t-x'/c) \nonumber\\
\end{eqnarray}
such that~\cite{Carmichael}
\begin{eqnarray*}
{\bf E}^{(+)}(x',t)={\rm i}\sum_j\hat{e}_j{\rm e}^{{\rm
i}\phi_{Tj}}\sqrt{\frac{\hbar\omega_{j}}{2\epsilon_0Ac}}
Q_j^{(+)}(x',t)~,
\end{eqnarray*}
for $ct>x'>0$. The decomposition of Eq.~(\ref{Rescale}) shows
how the photons transmitted through
the mirror into the cavity output mix with the external fields reflected by the mirror itself.

Let us now discuss the dynamics of the cavity field, assuming that at $t=0$
a pulse of duration $T$ is applied which fulfills~(\ref{Condition}).
At times $t > T$ the field inside the cavity evolves
according to
\begin{eqnarray}
\label{a:T}
a_j(T+t)&\approx& a_j(T){\rm
e}^{-({\rm i}\omega_j+\kappa_j)t}\\
&-&\sqrt{\frac{c}{L'}}\sqrt{2\kappa_j}
\int_0^t{\rm d}\tau{\rm e}^{-({\rm i}\omega_j+\kappa_j)(t-\tau)}
r_f^{(j)}(\tau)~.\nonumber
\end{eqnarray}
For later convenience, we generalize definition~(\ref{Rescale})
and consider the rescaled field operator
\begin{eqnarray}
Q_j(x',t,\theta_j) &=&Q_j^{(+)}(x',t){\rm e}^{{\rm i}\theta_j}+
Q_j^{(-)}(x',t){\rm e}^{-{\rm i}\theta_j}\nonumber\\
&=&{\cal Q}_0(t)(q_j(\theta_j)+\delta q_j(t,\theta_j))~,
\end{eqnarray}
where ${\cal Q}_0(t)=\sqrt{2\kappa_j}{\rm e}^{-\kappa_jt}$ is a
time-dependent scalar, $q_j(\theta_j)$ is the cavity-field
quadrature,
$$q_j(\theta_j)= a_j(T){\rm e}^{{\rm i}\theta_j}+a_j^{\dagger}(T){\rm e}^{-{\rm i}\theta_j}~,$$
and $\delta q_j(t,\theta_j)$ is the correspondingly defined
quadrature of the free field.

\subsection{Correlations in the fields at the cavity output}
\label{Sec:Entanglement}

Quantum correlations in the two-mode output
field ${\bf E}_{\rm out}$ are detected by balanced
homodyne detectors~\cite{Reid}, using local oscillators
$E_1^{(LO)}$, $E_2^{(LO)}$ with phases $\theta_1$ and $\theta_2$,
respectively, which mix with the fields previously spatially separated
by a beam splitter. A possible implementation is described in the following section.
The measured currents at the detectors are $i_1(t)= \alpha
Q_1(\theta_1)$ and $i_2(t)= \alpha Q_2(\theta_2)$, where $\alpha$
is a scaling parameter assumed to be equal for the two modes. The
correlations are measured through the combined difference current
$i_-(t)=i_1-i_2$, with
$$i_-(t)=\alpha\left(Q_{1}(t,\theta_1)- Q_2(t,\theta_2) \right)~.$$
We evaluate the current fluctuations at time $t$ on a grid $\delta
t$, such that $\kappa\delta t\ll 1$, i.e. fluctuations are
recorded on a time scale much faster than the cavity decay time.
The fluctuations of the difference current are given by
\begin{eqnarray}
\langle i_-(t)^2 \rangle &=& i_-^{(0)} C_{1,2}(t) \label{i:-}
\end{eqnarray}
where
$$i_-^{(0)}=\alpha^2 \left( \langle
Q_1(t,\theta_1)^2 \rangle + \langle Q_2(t,\theta_2)^2 \rangle
\right)
$$
is a positive proportionality constant, and
\begin{equation}
\label{C:XY} C_{1,2}(t)=1-\frac{2\langle Q_1(t,\theta_1)
Q_2(t,\theta_2)\rangle}{\langle Q_1(t,\theta_1)^2\rangle + \langle
Q_2(t,\theta_2)^2\rangle}
\end{equation}
contains the effect of quantum correlations. In these equations
the mean value $\langle\cdot\rangle$ of the operators at time $t$
is averaged over the interval of time $\delta t$ and the average
is taken over the vacuum state of the electromagnetic field. Using
Eqs.~(\ref{E:s}),~(\ref{E:f}) and~(\ref{a:T}), term~(\ref{C:XY})
in the difference current~(\ref{i:-}) takes the form
\begin{eqnarray}
C_{1,2}(t)=1-\frac{{\cal R}(t)}{1+{\cal R}(t)}\frac{ \langle q_{1}
(\theta_1)q_{2}(\theta_2)\rangle}{\langle
q_{1}(\theta_1)^{2}\rangle + \langle q_{2}(\theta_2)^{2}\rangle}~.
\label{C:12:2}
\end{eqnarray}
In Eq.~(\ref{C:12:2}) the relevant quantities are
\begin{equation}
\label{R(t)} {\cal R}(t)=\kappa\delta t{\rm e}^{-2\kappa
t}\left(\langle q_{1}(t,\theta_1)^2\rangle + \langle
q_{2}(t,\theta_2)^{2}\rangle\right)~, \end{equation} and
\begin{eqnarray}
\label{q:1} &&\langle q_{1}(\theta_1)^{2}\rangle=\langle
q_{2}(\theta_2)^{2}\rangle=
\frac{(|\chi_1|^2+|\chi_2|^2)^2+4|\chi_1\chi_2|^2}{\Theta^4}~, \nonumber\\
&& \\
&&\langle q_{1}(\theta_1)q_{2}(\theta_2) \rangle={\rm
Re}\left\{\frac{4\chi_1\chi_2(|\chi_1|^2+|\chi_2|^2)}{\Theta^4}{\rm
e}^{{\rm i}(\theta_1+\theta_2)}\right\}~. \nonumber\\
&&\label{q:12}
\end{eqnarray}
The details of the derivation of this result are reported in
appendix A.

Let us now discuss function $C_{1,2}(t)$,
Eq.~(\ref{C:12:2}), in detail. The second term on the rhs of
Eq.~(\ref{C:12:2}) is proportional to $\langle q_{1}
(\theta_1)q_{2}(\theta_2)\rangle$ and gives the effect of quantum
correlations. In absence of correlation between the two modes the average
$\langle q_{1} (\theta_1)q_{2}(\theta_2)\rangle$ vanishes and
$C_{1,2}(t)=1$. This value is the shot noise limit for independent
vacuum inputs into the homodyne detectors.

The correlations $\delta(X_1-X_2)^2$ and $\delta(P_1+P_2)^2$ of
the orthogonal quadratures $X$ and $P$ are obtained by setting
$\theta_1=\theta_2=0$ and $\theta_1=-\theta_2=\pi/2$,
respectively, which leads to identical results, namely
\begin{equation}
\label{C:12:r}
C_{1,2}(t)=1-\frac{{\cal R}(t)}{1+{\cal R}(t)}\frac{2r}{1+r^2}\frac{2}{1+\left(2r/(1+r^2)\right)^2}~,
\end{equation}
where we have used definition ~(\ref{r}). Thus, the regime
$C_{1,2}(t)<1$ corresponds to detecting EPR-type
entanglement~\cite{Reid,Giacobino,Giedke03}. Indeed, in this
regime the value of $C_{1,2}(t)$ is an entanglement
measure~\cite{Giedke03}. The effect on $C_{1,2}(t)$ of vacuum fluctuations that mix with the quantum
correlations of the cavity field at the cavity output is
represented by the parameter ${\cal R}(t)$, Eq.~(\ref{R(t)}). This
value is proportional to the number of photons inside the
cavity, and goes to zero as a function of time on a scale
determined by cavity decay. Therefore, for short times and large
number of photons the effect of quantum correlations in the cavity
field is well visible over the quantum noise. As the intensity of
the source field diminishes with time, the signal reaches the shot
noise limit.

\begin{center}
\begin{figure}
\epsfig{width=0.8\hsize, file=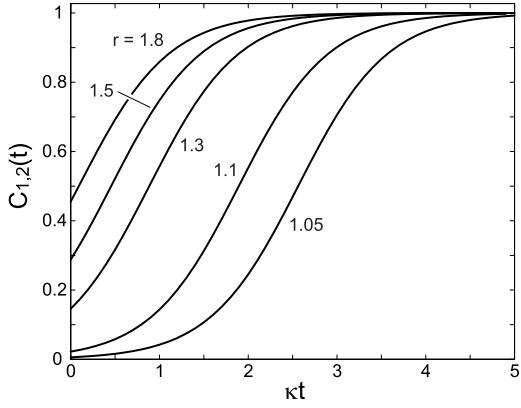} \caption{Signal
$C_{1,2}(t)$ as a function of time, for $\theta_1=\theta_2$ and
for values of the parameter $r=1.8,1.5,1.3,1.1,1.05$ (from left to
right). A time resolution of $\delta t = 0.1/\kappa$ is used. The
other parameters are discussed in the text.} \label{Fig:2}
\end{figure}
\end{center}

Figure~\ref{Fig:2} shows the signal $C_{1,2}(t)$ for different values of
the parameter~$r$. A reduction below 10\% of the shot
noise level is reached on a time scale of $1/\kappa$ for $r=1.1$.
About 110 photons per mode are created in this case. It should be
noted that for $r$ close to 1, significant two-mode squeezing is
observed over several cavity decay times, before the shot noise
level $C_{1,2}(t)=1$ is approached. This occurs when the number of
photons remaining in the cavity reaches the order of one.

\subsection{Experimental parameters}
\label{Sec:Experiment}

We now discuss the parameter regime in which our description
holds. First let us consider the ideal dynamics, as given by
Eqs.~(\ref{Heisen:1})-(\ref{Heisen:3}). The degree of squeezing in
the two-mode state~(\ref{Two:mode}) is fixed by the ratio
$r=|\chi_2/\chi_1|$, Eq.~(\ref{r}); large squeezing requires $r$
be close to 1. With the cavity mode frequencies much larger than
the trap frequency, we can assume $g_1 \approx g_2 = g$.
Therefore, the degree of squeezing is solely controlled by the
quantum effects in the mechanical action of the light, which enters
through the ratio $\nu/\Delta$ between the trap frequency and the
detuning (in this section we assume for convenience $\Delta$ for $|\Delta|$).
A small value of this ratio, i.e.\ $\nu \ll \Delta$,
means large two-mode squeezing. Indeed, the control of the degree
of squeezing through this ratio implies a further condition
relating the linewidth of the transition $\gamma$ and the trap
frequency $\nu$, namely
\begin{equation}
\label{Condition:0} \nu\gg\gamma~,
\end{equation}
such that $\Delta\gg\nu\gg\gamma$.
Under these conditions we find
\begin{equation}
\label{r:1} r \approx 1+2\frac{\nu}{\Delta}~.
\end{equation}
Furthermore we recall that $\Delta\gg\Omega,g$ was required throughout the model.

We now derive further conditions under which the dynamics are
described by Eqs.~(\ref{Heisen:1})-(\ref{Heisen:3}). We have
already identified in Eq.~(\ref{Condition}) an upper and a lower
bound to the duration $T$ of the laser excitation pulse, due to
cavity decay and to the spectral resolution of the cavity modes
(whose frequency separation is fixed to twice the trap frequency).
Other restrictions result from the requirement that processes in
which the atom scatters laser photons into modes external to the
cavity are negligible. Here some distinctions must be made. In
fact, elastic scattering and inelastic scattering along the
carrier (i.e.\ without changing the motion) do not affect the
relevant dynamics, since they do neither change the number of
phonons or cavity photons nor dephase their quantum states.
Detrimental processes are (i) inelastic scattering of laser
photons along the sidebands (i.e.\ changing the motion) and (ii)
scattering of cavity photons into the external modes. Processes of
type (i), which would add dissipation to Eq.~(\ref{Heisen:2}), are
characterized by a rate $\gamma_{\Theta} \sim
\gamma\eta^2\Omega^2/\Delta^2$. During an excitation pulse of
duration $T \sim 1/\Theta$ they are negligible as long as
\begin{equation}
\label{Condition:2} \gamma_{\Theta}  \ll\Theta~.
\end{equation}
Processes of type (ii) occur at a rate $\gamma_{\kappa}\sim \gamma
g^2/\Delta^2$. They enter as dissipative terms into
Eqs.~(\ref{Heisen:1}) and (\ref{Heisen:3}), and are negligible
provided that
\begin{equation}
\label{Condition:3} \gamma_{\kappa}\ll\kappa.
\end{equation}
Moreover, the coherent dynamics are based on the validity of the
Lamb-Dicke regime at all times $0 \le t \le T$. This corresponds
to the condition $\eta\sqrt{\langle b^{\dagger}(t)b(t)\rangle} \ll
1$ which can be rewritten as
\begin{equation}
\label{Condition:4} \eta\sqrt{\Delta/4\nu}\ll 1
\end{equation}
using Eqs.~(\ref{b}),~(\ref{r}) and~(\ref{r:1}). Finally,
in the case of ion traps
decoherence of the center-of-mass oscillation can safely be
ignored, as the trapping potential has been experimentally
demonstrated to be very stable on time scales of the order of
milliseconds~\cite{Trap:Stability}.

We now identify parameter regimes where significant shot noise
reduction can be reached while conditions (\ref{Condition}),
(\ref{Condition:2}), (\ref{Condition:3}), and (\ref{Condition:4})
are simultaneously fulfilled. We use~\cite{Kimble-lecture}
$$%
g = \sqrt{\frac{\sigma}{4\pi A}} \sqrt{\gamma {\rm \delta\omega}}
\equiv \sqrt{\tilde{\sigma}} \sqrt{\gamma {\rm \delta\omega}}
$$%
where $\sigma \propto \lambda^2$ is the scattering cross section
of the atom in free space, $A$ is the cavity mode waist, $L$ is
the cavity length and ${\delta\omega}=2\pi c/2L$ is the cavity
free-spectral range. The cavity decay rate is
$\kappa=\frac{\delta\omega}{\cal F}$ where $\cal F$ is the
finesse. The condition $1/T\gg\kappa$ in Eq.~(\ref{Condition}) together with $T=\pi/\Theta$
imposes the relation
\begin{eqnarray}
\frac{\Theta}{\kappa}
 &=& \sqrt{2}\eta\sqrt{\frac{2\nu}{\Delta}}
     \frac{\Omega}{\Delta}\frac{g}{\kappa} \nonumber \\
 &=& {\cal F} \sqrt{\frac{4\nu\tilde{\sigma}}{\Delta}}
     \left( \eta\frac{\Omega}{\Delta}
     \sqrt{\frac{\gamma}{\delta\omega}} \right) \gg 1
\label{a:Condition}
\end{eqnarray}
where we have used Eqs.~(\ref{Chi:1}) and~(\ref{Chi:2}) taking
$\cos\theta_L=1$ and $\cos\theta_c=0$, i.e., the laser wave
vector parallel to the motional axis, and the cavity wave vector
perpendicular to both. Condition~(\ref{Condition:2}) leads to the
relation
\begin{eqnarray}
\frac{\Theta}{\gamma_{\Theta}}
 &=& \sqrt{2}\sqrt{\frac{2\nu}{\Delta}} \left(
     \eta\frac{\Omega}{\Delta} \right)^{-1} \frac{g}{\gamma}
     \nonumber\\
 &=& \sqrt{\frac{4\nu\tilde{\sigma}}{\Delta}}
     \left( \eta\frac{\Omega}{\Delta}
     \sqrt{\frac{\gamma}{\delta\omega}} \right)^{-1} \gg 1~.
\label{a:Condition:2}
\end{eqnarray}
Finally, from (\ref{Condition:3}) we find
\begin{eqnarray}
\frac{\kappa}{\gamma_{\kappa}}
=\frac{\Delta^2}{\gamma^2}\frac{1}{{\cal F} \tilde{\sigma}} \gg
1~. \label{a:Condition:3}
\end{eqnarray}
These inequalities can conveniently be summarized as
\begin{equation}
 \frac{4\Delta\nu}{\gamma^2} \gg
 {\cal F} \frac{4\nu\tilde{\sigma}}{\Delta}  \gg
 \sqrt{\frac{4\nu\tilde{\sigma}}{\Delta}} \left(
 \eta\frac{\Omega}{\Delta} \sqrt{\frac{\gamma}{\delta\omega}}
 \right)^{-1} \gg 1~.
\label{Supercondition}
\end{equation}

We consider now the ratio~$\nu/\Delta=0.05$, which gives $r=1.1$
(Eq.~(\ref{r:1})), corresponding to significantly squeezed pulses
with an average number of about 110 photons per mode. Taking
realistic values $\eta=0.1$, $\Omega/\Delta=0.3$, and
$\tilde{\sigma}= 10^{-3}$, we obtain
\begin{equation}
80 \left( \frac{\nu}{\gamma} \right)^2 \gg 2 \times 10^{-4}{\cal
F} \gg\ 0.5 \sqrt{\frac{\delta\omega}{\gamma}} \gg 1~.
\label{Eq:Supernumbers}
\end{equation}
Additionally, Eq.~(\ref{Condition:0}) is required;
condition~(\ref{Condition:4}) is already met with the given choice
of the Lamb-Dicke parameter $\eta$ and of the ratio $\nu/\Delta$.

A possible system to fulfil Eq.~(\ref{Eq:Supernumbers}) and thus
implement the desired dynamics is a single In$^+$ ion
\cite{Peik1999}, confined by an ion trap of frequency
$\nu=2\pi\times 3$~MHz, laser-excited on its intercombination line
at 231~nm (linewidth $\gamma=2\pi\times 360$~kHz) at
$\Delta=2\pi\times 60$~MHz detuning and $\Omega=2\pi\times 18$~MHz
Rabi frequency, and coupled to an optical cavity with free
spectral range ${\delta\omega}=2\pi\times 1$~GHz and finesse
${\cal F}=10^6$. For these parameters, $g \approx 2\pi\times
0.6$~MHz, $\Theta \approx 2\pi\times 8$~kHz and $\kappa \approx
2\pi\times 1$~kHz, and one would measure highly entangled pulses,
characterized by 99\% reduction of the vacuum fluctuations over a
time of the order of 0.1 msec.

\begin{center}
\begin{figure}
\epsfig{width=0.95\hsize, file=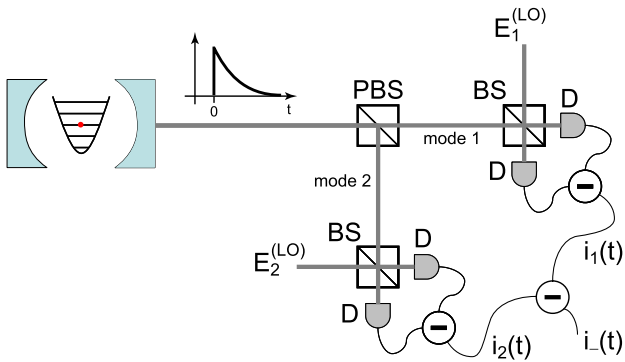}
  \caption{Schematic set-up for measuring quantum correlations
  in the field at the cavity output. PBS stands for
  polarizing beam splitter, BS for beam splitter,
  D for detector. Details of the experimental set-up are discussed in the text.}
  \label{Fig:setup}
\end{figure}
\end{center}

%A schematic layout of an experimental set-up to realize our model
%is shown in Fig.\ref{setup}. An optical cavity, with parameters as
%given above, is built around an atom trap (or ion trap) containing
%a single atom. The atom is driven by a laser field which
%propagates orthogonally to the cavity axis.
To obtain a frequency splitting of $2\nu \ll \delta\omega$ between
the two cavity modes involved in the dynamics, two non-degenerate
polarization modes may be utilized, both of which couple to the
laser-driven transition of the atom. With In$^+$, this is achieved
by setting the quantization axis $\vec{B}$ along the cavity axis,
and $\vec{B}$, $\vec{k}$, and laser polarization $\vec{E}_L$
mutually orthogonal. Other possible atomic level schemes are,
e.g., a ${J}\!=\!1/2 \leftrightarrow {J^{\prime}}\!=\!1/2$ or an
${F}\!=0 \leftrightarrow {F^{\prime}}\!=\!1$ transition. The
two-mode field emitted from the cavity after the laser excitation
pulse is split by a polarizing beam splitter, and the fluctuations
of both modes are detected by balanced homodyne detectors, as
shown schematically in Fig.~\ref{Fig:setup}.

\section{Temporally separated entangled pulses}
\label{Sec:Time_separation}

We now discuss a scheme which allows for the creation of pairs of
temporally separated, entangled pulses, which may be
monochromatic. The scheme is based on an atom trapped in a cavity,
of which only one mode is relevant to the dynamics and that is far
off-resonance from the dipole transition. The scheme is connected
to ideas of quantum state transfer between the quantum
center-of-mass motion and the electromagnetic
field~\cite{Ze-Parkins99}, and to its possible applications for
creating quantum correlations between distant
atoms~\cite{Parkins02}. In the present case, the ion keeps the
memory of the quantum correlations with the field which is emitted
by the cavity, and transfers it to the subsequent pulse. The
resulting pulse pair exhibits EPR-type correlations in the quantum
fluctuations.

\subsection{Dynamics}

We denote by $a,a^{\dagger}$ the annihilation and creation
operator of a cavity photon at frequency $\omega_c$, and assume
that both the center-of-mass and the cavity oscillator are
initially prepared in the vacuum state. The model description is
analogous to the one given in Sec.~\ref{Sec:Theory}, whereby now
the sum over the cavity modes is dropped, together with the
subscript $j$.

The generation of pairs of monochromatic entangled pulses follows
this procedure: first, a laser pulse is applied with
$\omega_L=\omega_c+\nu$, in the regime in which the relevant
dynamics are described by Hamiltonian $H_{\rm
eff}^{(1)}=\hbar\omega_c a^{\dagger}a+H_{\rm mec}+H^{(1)}$ with
\begin{eqnarray}
H^{(1)}={\rm i}\hbar\chi a^{\dagger}b^{\dagger}+{\rm
 H.c}\label{H:1:1}
\end{eqnarray}
where the coupling term
\begin{eqnarray}
\chi=\eta \frac{g^*\cos\phi\Omega(t)}{\Delta}\left(\cos\theta_L
+{\rm i}\tan\phi\cos\theta_c\right)\label{Chi:1:1}
\end{eqnarray}
has been obtained in the limit $|\Delta|\gg\nu,\gamma$. Therefore, at the
end of the laser pulse atomic motion and cavity mode are two-mode squeezed,
their degree of squeezing being determined by the duration $T_1$
of the pulse according to
\begin{eqnarray*}
&&a(T_1)=a(0)\cosh|\chi|T_1+b^{\dagger}(0){\rm e}^{{\rm
i}\phi_{\chi}}\sinh|\chi|T_1\\
&&b^{\dagger}(T_1)=b^{\dagger}(0)\cosh|\chi|T_1+a(0){\rm e}^{-{\rm
i}\phi_{\chi}}\sinh|\chi|T_1~,
\end{eqnarray*} with $\chi={\rm
e}^{{\rm i}\phi_{\chi}}|\chi|$. The average occupation number of
both the cavity mode and the center-of-mass oscillator is $\langle
n\rangle=\sinh^2|\chi| T_1$. The cavity field evolves according to
Eq.~(\ref{a:T}) and after several decay times it is in the vacuum
state, while the field at the cavity output is characterized by a
propagating pulse described by Eq.~(\ref{E:s}), whose amplitude
fluctuations are entangled with the motional state.

Let us then assume that at time $\tau\gg 1/\kappa$ a second laser
pulse is applied, which is now tuned  to $\omega_L=\omega_c-\nu$,
thus driving dynamics described by Hamiltonian $H_{\rm
eff}^{(2)}=\hbar\omega_c a^{\dagger}a+H_{\rm mec}+H^{(2)}$ with
\begin{eqnarray}
H^{(2)}={\rm i}\hbar\chi a^{\dagger}b+{\rm
 H.c}\label{H:2:1}
\end{eqnarray}
at the same coupling constant $\chi$ as in Eq.~(\ref{Chi:1:1}).
For a pulse duration $T_2=\pi/2|\chi|$ then
\begin{eqnarray*}
a(\tau+T_2)=b(\tau){\rm e}^{{\rm i}\phi_{\chi}}
\end{eqnarray*}
and the motion is in the state of the cavity field at time $\tau$.
In absence of decoherence processes for the atomic motion, then
$b(\tau)=b(T_1)$.
Therefore, for $\kappa\tau\gg 1$, at the end of the second laser
pulse the motion becomes decorrelated from the first propagating
pulse and its correlations have been transferred to the cavity
field, which is in a two-mode squeezed state with the first
propagating pulse.

\subsection{Field at the cavity output}

The cavity field, entering Eq.~(\ref{E:s}) as the source field, is
written as
\begin{eqnarray}
\nonumber
a(t)&\approx& \theta(t-T_1)a(T_1){\rm e}^{-\kappa t}\nonumber\\
      & +    &\theta(t-\tau-T_2)a(\tau+T_2)
{\rm e}^{-\kappa (t-\tau)}\nonumber\\
&-&\sqrt{\frac{c}{L'}}\sqrt{2\kappa}
\int_0^t{\rm d}\tau{\rm e}^{-({\rm i}\omega+\kappa)(t-\tau)}
r_f(\tau)\label{a:T:1}
\end{eqnarray}
and the source field~(\ref{E:s}) describes now two temporally
separated pulses, whose separation can be controlled on a time
scale of the order of the cavity decay time.

The correlation between the pulses can be detected by measuring
the fluctuations of the difference current between the signals at
the detector at $t$ and $t+\tau$, which we define as
\begin{equation}
\tilde{i}_-(t,\tau)=\alpha(Q(t,\theta_1)-Q(t+\tau,\theta_2))
\end{equation}
The correlation function show the same functional behaviour as
function~(\ref{C:12:r}) where now $r$ is related to $|\chi|$ and
$T_1$ by $\tanh |\chi|T_1=2r/(1+r^2)$.

\subsection{Experimental parameters}

This type of proposal requires the coupling of the dipole with a
single cavity mode, and it therefore simplifies several
experimental conditions with respect to the simultaneous
generation of bichromatic entangled pulses, see Sec.~\ref{Sec:II}.
We list below some salient requirements.

Coherent dynamics during the laser pulse is achieved provided that
$T_1,T_2\ll 1/\kappa$. Moreover, spectral resolution of the
vibrational excitations imposes $T_1,T_2\gg 1/\nu$. Therefore, an
important condition for the realization of this scheme is
\begin{equation}
\label{Condition:T} \nu \gg \frac{1}{T_1}, \frac{1}{T_2} \gg
\kappa~.
\end{equation}
This condition is accompanied by the requirements on negligible
incoherent scattering by the atom, $\gamma_{\kappa} \ll \kappa$
(Eq.~(\ref{Condition:3})) and $\gamma_{\Theta}T_1\ll 1$, which is
equivalent to condition~(\ref{Condition:2}) for this type of
scheme. Moreover, the Lamb-Dicke regime must be fulfilled at any
stage of the dynamics. Therefore, large reductions in photon
number correlations below the shot noise limit can be produced
with atoms confined in very tight traps, i.e.\ with very small
Lamb-Dicke parameters. This in turn affects the speed of the
dynamics, as the coupling $\chi$ scales with $\eta$. For instance,
after the first pulse the average number of vibrational
excitations (and of cavity-mode photons) is $\langle
n\rangle=\sinh^2|\chi| T_1$, hence the Lamb-Dicke regime is
fulfilled at all stages provided that
\begin{equation}
\label{Condition:T:1} \eta \Bigl|\sinh|\chi| T_1\Bigr|\ll 1
\end{equation}
If we set $\langle n\rangle \approx 100$, thus imposing
$\eta=0.03$, then $T_1\approx \log 20/|\chi|$. Taking these
values, $\Omega/\Delta=0.3$, and $T_2 \approx T_1$, we find from
condition~(\ref{Condition:T}) a relation for the cavity parameters
and the trap frequency,
\begin{equation}
\nu \gg g/300 \gg \kappa~.
\end{equation}
Besides this, there is no particular requirement on the ratio
$\gamma/\nu$, therefore these numbers can be obtained with various
atomic species in experimentally available set-ups. It should be
noticed that reliable entanglement between the temporally
separated pulses requires that the coherence of the quantum state
of the center-of-mass oscillator is preserved during and between
the pulses. Generally, for ion traps one can rely on heating times
of the order of tens of milliseconds, such that this condition is
fulfilled~\cite{Trap:Stability}. A study of decoherence on the
efficiency of the scheme will be the subject of future work.

Finally, we comment on the initial preparation of the
center-of-mass state. The dynamics discussed here apply when the
motion is prepared in the ground state of the confining potential,
which may be achieved with ground state cooling techniques.
However, initial preparation of the center-of-mass oscillator in
the vacuum state is also possible by means of quantum state
transfer techniques between the motion and the electromagnetic
field~\cite{Ze-Parkins99}. Since these techniques are at the basis
of the entanglement scheme, ground state cooling is not a
necessary requirement. In future studies we will also investigate
the scheme when the motion has been prepared in a different state
than the ground state.

\section{Conclusions}
\label{Sec:Conclusions}

To conclude, we have shown that the motion of a single trapped
atom inside an optical cavity can act as a quantum medium which
mediates entanglement on demand between simultaneous or subsequent
radiation pulses. The process is based on the mechanical effect of
light, which in the quantum regime allows for coherently
controlling the interaction and thereby the degree of
entanglement. We have discussed two schemes, which allow for simultaneous bichromatic
and temporally separated entangled pulses. From our estimates the proposal requires
experimental regimes that are within reach, and would allow for
the production of entangled light pulses on demand, characterized
by 99\% reduction of the vacuum fluctuations over a time of the
order of 0.1 msec.

Our schemes offer interesting alternatives to
implementations with atomic gases~\cite{Kimble03, Polzik04}, where
now the controlled interaction with the spectrum of the quantum
excitations creates the entanglement with the radiation pulses.
It can be extended to the microwave regime by
suitably driving atomic microwave transitions in a setup like the
one discussed in~\cite{Wunderlich}. It can also be extended to the
collective excitations of ultracold atomic gases, where the nature
of the \textit{collective} excitations would allow for additional
freedom in tuning the parameters, thereby giving rise to higher
efficiencies or new properties of the emitted radiation.
The scheme with spatially separated entangled pulses
may be of help in devising new
cryptographic schemes exploiting time correlated pulses
and continuous alphabets,
thus extending those based on time-energy entangled photon pairs
(see e.g. \cite{Gisetal02} and references therein).

In the future we will study correlations in the continuous-wave
excitation of the ion, in the perspective of applications for
quantum networking, like for instance discussed
in~\cite{CiracKimble, Kraus04}.

\begin{acknowledgments}
The authors gratefully acknowledge discussions with Christoph
Becher, Luiz Davidovich, Markus Hennrich, Paulo Nussenzveig, and
Scott Parkins. This work was partly supported by the European
Commission (CONQUEST network, MRTN-CT-2003-505089; SCALA, Contract No.\ 015714), and the
scientific exchange programme Spain-Italy (HI2003-0075); G.~M. is
supported by the Spanish Ministerio de Educaci\'on y Ciencia
(Ramon-y-Cajal).
\end{acknowledgments}

\begin{appendix}

\section{Evaluation of the field correlation functions}

In this appendix we report the detailed steps for the explicit
derivation of the difference current~(\ref{i:-}). We assume
$\kappa_1\approx\kappa_2$. The single terms on the right-hand side
of Eq.~(\ref{i:-}) are evaluated to be
\begin{eqnarray}
&&\langle Q_1(t,\theta_1)^2\rangle = 2\kappa {\rm e}^{-2\kappa t}
\langle q_1(\theta_1)^{2} \rangle + \frac{c}{L'}\sum_k
\frac{\omega_{k}^{(1)}}{\omega_1}{\cal I}_1(t)
\nonumber\\
&&\label{X2} \\
&&\langle Q_2(t,\theta_2)^2\rangle = 2\kappa {\rm e}^{-2\kappa t}
\langle q_2(\theta_2)^{2}\rangle + \frac{c}{L'}\sum_k
\frac{\omega_{k}^{(2)}}{\omega_2} {\cal I}_2(t)
\nonumber\\
&&\label{Y2} \\
&& \langle Q_1(t,\theta_1) Q_2(t,\theta_2)\rangle = 2\kappa
{\rm e}^{-2\kappa t}\langle q_1(\theta_1) q_2(\theta_2) \rangle \nonumber\\
\label{C:XY:0}
\end{eqnarray}
where
\begin{eqnarray}
\langle q_1(\theta_1)^{2}\rangle %
&=& \langle (a_{1}{\rm e}^{{\rm i}\theta_1}+ a_{1}^{\dagger}{\rm e}^{-{\rm i}\theta_1})^2 \rangle \nonumber\\
&=& \frac{(|\chi_1|^2+|\chi_2|^2)^2+4|\chi_1|^2|\chi_2|^2}{\Theta^4}~, \\
\langle q_2(\theta_2)^{2}\rangle &=& \langle
q_1(\theta_1)^{2}\rangle~,
\end{eqnarray}
and
\begin{eqnarray}
&&\langle q_1(\theta_1) q_2(\theta_2) \rangle={\rm
Re}\left\{\frac{4\chi_1\chi_2(|\chi_1|^2+|\chi_2|^2)}{\Theta^4}{\rm
e}^{{\rm i}(\theta_1+\theta_2)}\right\}~, \nonumber\\
\end{eqnarray}
while the integral
\begin{eqnarray}
{\cal I}_j(t) &=&\frac{1}{\delta t^2}\int_t^{t+\delta t} {\rm
d}\tau\int_{t}^{t+\delta t} {\rm d}\tau'{\rm e}^{-{\rm
i}\omega_{k}^{(j)}(\tau-\tau')}\nonumber\\
&=&2\frac{\sin\omega_{k}^{(j)}\delta t}{\omega_{k}^{(j)}\delta t}
\label{Int} \end{eqnarray} introduces the finite spectral
resolution associated with the temporal grid. With Eq.~(\ref{Int})
in Eqs.~(\ref{X2}) and~(\ref{Y2}), we rewrite the sum over the
free field modes as
\begin{eqnarray*}
\sum_k\frac{\omega_{k}^{(j)}}{\omega_j}{\cal I}_j(t) &\sim&
\frac{1}{\delta t}~,
\end{eqnarray*}
where we have taken the continuum limit of the discrete sum over
the modes, thereby adding the density of states and assuming that
$\omega_{k}^{(j)}$ varies negligibly over $1/\delta t$.
Substituting into Eqs.~(\ref{X2}),~(\ref{Y2}), we obtain
\begin{eqnarray}
&&\langle Q_1(t,\theta_1)^2\rangle=2\kappa{\rm e}^{-2\kappa
t}\left(\langle q_1(\theta_1)^{2}\rangle +\frac{{\rm e}^{2\kappa
t}}{2\kappa \delta t}\right)~,
\label{Q:1}\\
&&\langle Q_2(t,\theta_2)^2\rangle=2\kappa {\rm e}^{-2\kappa
t}\left(\langle q_2(\theta_2)^{2}\rangle +\frac{{\rm e}^{2\kappa
t}}{2\kappa\delta t}\right) \label{Q:2}~.
\end{eqnarray}
Taking $\omega_1\approx \omega_2$ we finally obtain
\begin{equation}
\label{C} C_{1,2}(t)=1-\frac{{\cal
R}(t)}{1+{\cal R}(t)}c_{1,2}(\theta_1,\theta_2)~,
\end{equation}
where ${\mathcal R}(t)$ is defined in Eq.~(\ref{R(t)}) and
\begin{eqnarray}
&&c_{1,2}(\theta_1,\theta_2)= \frac{2{\rm Re}\{\langle
q_1(\theta_1) q_2(\theta_2)\rangle\}}{\langle
q_1(\theta_1)^{2}\rangle + \langle q_2(\theta_2)^{2}\rangle}.
\end{eqnarray}

\end{appendix}

\end{document}